\documentclass[10pt,conference]{IEEEtran}
\IEEEoverridecommandlockouts

\usepackage{amsmath,amssymb,amsfonts}
\usepackage{amsthm}
\usepackage{cite}
\usepackage{adjustbox}

\usepackage{booktabs}
\usepackage{array}
\usepackage[ruled]{algorithm2e}
\usepackage{algorithmic}

\newtheorem{remark}{Remark}
\usepackage{graphicx}
\usepackage[caption=false,font=footnotesize]{subfig}
\usepackage{textcomp}

\usepackage{url}

\usepackage{enumitem}
\usepackage{xcolor}
\def\BibTeX{{\rm B\kern-.05em{\sc i\kern-.025em b}\kern-.08em
    T\kern-.1667em\lower.7ex\hbox{E}\kern-.125emX}}
\begin{document}

\title{\huge Energy Efficiency Maximization for Integrated Sensing and Communications in Satellite-UAV  MIMO Systems\\
}

\author{Ngo Tran Anh Thu, \textit{Student Member}, \textit{IEEE}, Pham Dang Anh Duc, Bui Trong Duc, Nguyen Minh Quan,\\ Trinh Van Chien, \textit{Member}, \textit{IEEE}, Hoang D. Le, \textit{Member}, \textit{IEEE}  
\thanks{The work of Hoang D. Le was supported by the Telecommunications Advancement Foundation (TAF). This research is funded by the Vietnam National Foundation for Science and Technology Development (NAFOSTED) under grant number 102.02-2025.73 for Trinh Van Chien.}
\thanks{Ngo Tran Anh Thu, Bui Trong Duc, Nguyen Minh Quan and Trinh Van Chien are with the School of Information and Communications Technology, Hanoi University of Science and Technology, Hanoi 100000, Vietnam (e-mail: anhthungo.tr@gmail.com, ducbt@soict.hust.edu.vn, chientv@soict.hust.edu.vn, quannm@soict.hust.edu.vn). Pham Dang Anh Duc is with the Phenikaa University, Hanoi 100000, Vietnam (email: duc.phamdanganh@phenikaa-uni.edu.vn), Hoang D. Le is with the Computer Communications Laboratory, the University of Aizu, Aizuwakamatsu 965-8580, Japan (email: hoangbkset@gmail.com).}
\thanks{\textit{Corresponding author: Trinh Van Chien}}
}


\maketitle

\begin{abstract}
This paper investigates energy efficiency maximization in an integrated sensing and communication framework for satellite-UAV MIMO systems, where a LEO satellite and a UAV simultaneously serve ground users and perform target sensing. Both the satellite and UAV are equipped with uniform planar arrays of transmit antennas, enabling a distributed multi-user and multi-target architecture. We derive the achievable downlink throughput by considering that the high-altitude satellite maintains a line-of-sight (LoS) link with users, while adopting a probabilistic model for the UAV that accounts for the likelihood of both LoS and non-line-of-sight conditions. The energy efficiency maximization problem is formulated as a complex non-convex optimization problem, subject to power constraints, quality of service (QoS) requirements, and beampattern gain constraints for accurate sensing. To tackle this challenge, we propose an efficient alternating optimization algorithm capable of handling the complex search space and QoS guarantees. Numerical results across diverse scenarios with multiple users demonstrate that the proposed method achieves high energy efficiency while meeting both communication and sensing performance targets.
\end{abstract}

\section{Introduction}
\label{sec:Intro}
Next-generation (NG) networks are envisioned to natively embed integrated sensing and communication (ISAC) as a core physical-layer functionality, where a unified waveform is jointly exploited for data transmission and environmental sensing~\cite{10979960,10681275}. By sharing spectrum, hardware, and signal processing resources, ISAC substantially enhances spectral efficiency, reduces power consumption, and lowers implementation cost, while enabling high-resolution tasks such as gesture sensing, environmental mapping, and real-time object tracking~\cite{10445319}.
However, terrestrial base station (BS)-centric ISAC architectures remain inherently limited by fixed infrastructure. Their performance deteriorates rapidly in the presence of severe NLoS blockages, sparse network deployments, disaster-prone conditions, and strong mutual interference in dense urban environments.
To overcome these bottlenecks, NG research is shifting toward UAV–satellite ISAC architectures, where space-air platforms complement the terrestrial network. Satellites provide resilient, wide-area connectivity independent of ground infrastructure, while UAVs offer highly agile LoS sensing and communication with adaptive positioning~\cite{10851844,10098686}. Therefore, their joint operation forms a flexible three-dimensional (3D) network that maintains service continuity in remote areas, emergency scenarios, and temporary hotspots, situations where terrestrial ISAC alone cannot guarantee reliability or coverage.

To the best of our knowledge, no existing work provides a comprehensive analysis of space–air ISAC systems that incorporates spatially correlated MIMO arrays and joint processing across satellite–UAV–terrestrial links. Although \cite{chen2024resilient} studies a satellite–UAV architecture, the satellite operates solely as a cloud processor and does not transmit. In contrast, we enable full MIMO-based satellite–UAV cooperation and quantify its energy-efficiency gains. Our main contributions are summarized as follows: $i)$ Propose a hybrid space–air ISAC architecture where satellites and UAVs jointly perform sensing and downlink transmission under spatially correlated Rician fading with imperfect CSI;
$ii)$ Derive downlink rate expressions and formulate a non-convex energy-efficiency maximization problem subject to QoS and beam-pattern constraints for joint terrestrial and aerial target sensing; $iii)$ Develop an alternating optimization framework that decomposes the problem into efficiently solvable subproblems, ensuring low complexity and fast convergence; and $iv)$ Validate the proposed framework numerically, demonstrating robust ISAC performance across diverse user distributions and system configurations.

\begin{figure*}[t]
    \centering
    \subfloat[]{\includegraphics[width=0.4\textwidth]{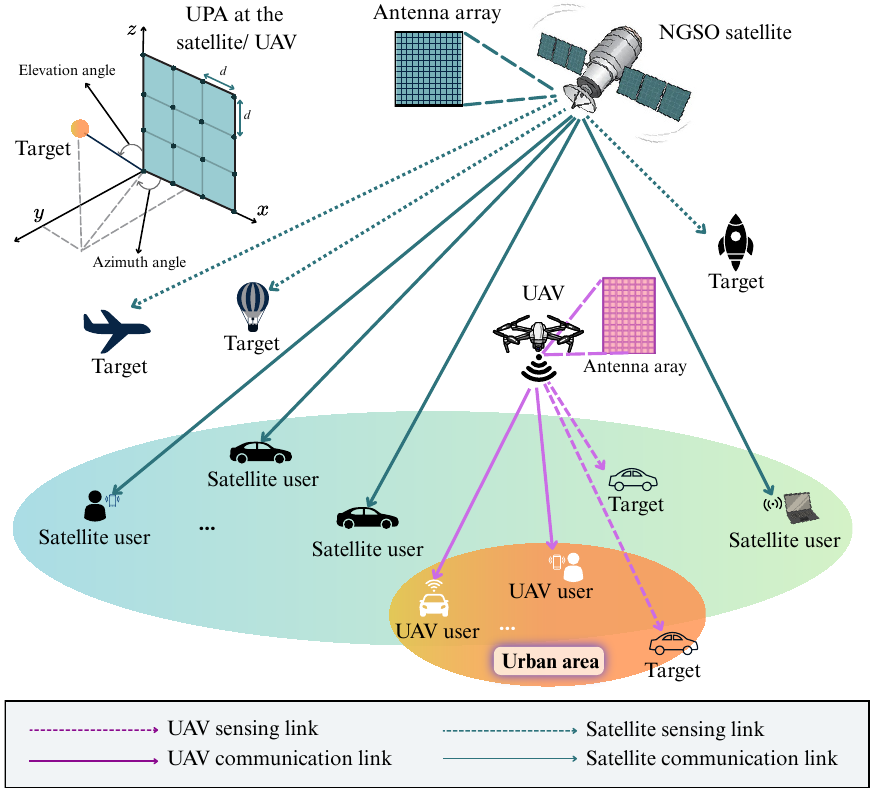}}\qquad
    \subfloat[]{\includegraphics[trim={0 0.5cm 0 0mm},clip,width=0.53\textwidth]{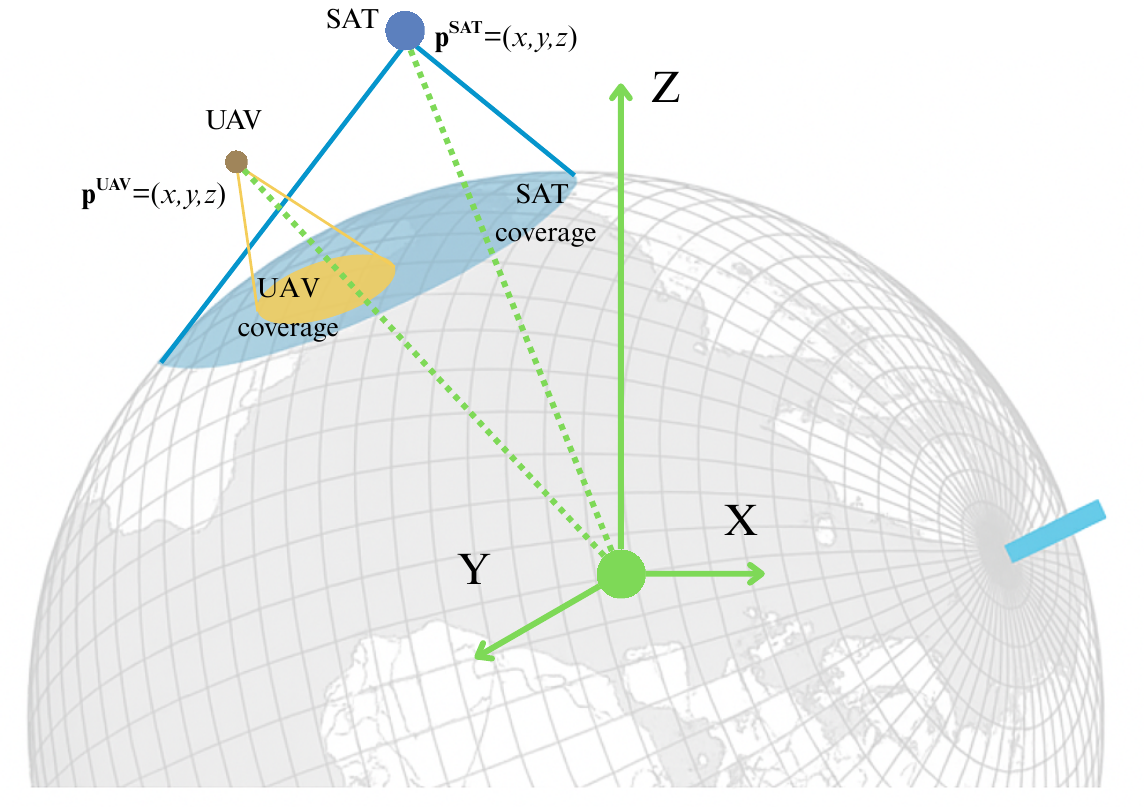}}
    \caption{Illustration of an ISAC Satellite-UAV wireless network: (a) Integrated space-aerial-terrestrial scenario with multi-functional sensing and communication links, (b) Geographical coverage of satellite-UAV platforms in a 3D coordinate system.}
    \label{fig:system_model}
\end{figure*}

\textit{Notation:} 
The matrix transpose and Hermitian transpose are denoted by $(\cdot)^T$ and $(\cdot)^H$, respectively. The Euclidean norm is represented by $|\cdot|$. $\mathcal{CN}(\cdot,\cdot)$ and $\mathcal{U}(\cdot,\cdot)$ denote complex Gaussian and uniform distributions, respectively. $\otimes$ represents the Kronecker product, $\odot$ denotes element-wise multiplication, and $j$ indicates imaginary unit. 

\section{Satellite-UAV-assisted Integrated Sensing and Communications Model}
\label{sec:space-aerial}

We consider a satellite–UAV ISAC system, as illustrated in Fig.~\ref{fig:system_model}. A LEO satellite and a UAV, indexed by $\xi \in \{sat, uav\}$, are equipped with uniform planar arrays (UPAs) of $M^{(\xi)} = M_x^{(\xi)} \times M_z^{(\xi)}$ transmit antennas. Each platform $\xi$ serves single-antenna downlink users $\mathcal{Q}^{(\xi)} = \{1, \ldots, N^{(\xi)}\}$ and senses targets $\mathcal{T}^{(\xi)} = \{1, \ldots, L^{(\xi)}\}$, where $N^{(\xi)}$ and $L^{(\xi)}$ denote the numbers of served users and sensing targets associated with platform $\xi$, respectively. Specifically, the satellite covers rural areas, while the UAV serves urban regions. Let $\mathbf{p}^{(\xi)}$ and $\mathbf{u}_k$ denote the positions of platform $\xi$ and ground user~$k$, respectively. The channel corresponds to a steering vector $\mathbf{a}^{(\xi)}_{k} \in \mathbb{C}^{M^{(\xi)} \times 1}$, parameterized by the elevation $\theta_k^{(\xi)}$ and azimuth $\varphi_k^{(\xi)}$ angles relative to the platform \cite{10851844}. In the downlink phase,  the satellite serves rural users with wide-area coverage, while the UAV covers urban regions due to its shorter range.
\subsection{Multi-user Downlink Communication Network}
\label{sec:DL Com}
In this paper, the satellite maintains a LoS link with ground users due to the satellite’s high altitude. In contrast, the UAV operates at a lower altitude, where both LoS and NLoS conditions may occur. 
We denote the vector $\mathbf{h}_{k} \in \mathbb{C}^{M^{(sat)}}$ as the downlink satellite-terrestrial channel in the Ka band between the LEO satellite and the user $k$, which can be expressed as 
\begin{equation}
\mathbf{h}_{k} = \pmb{\ell}_{k} \odot \left( 
\sqrt{ \frac{\mathcal{K}}{\mathcal{K} + 1} } \mathbf{h}^{\text{LOS}}_{k} 
+ \sqrt{ \frac{1}{\mathcal{K} + 1} } \mathbf{h}^{\text{NLOS}}_{k} 
\right),
\label{eq:hk_m}
\end{equation}
where $\mathcal{K}$ represents the Rician factor, $\mathbf{h}^{\text{LOS}}_{k}$ and $\mathbf{h}^{\text{NLOS}}_{k}$ are the LOS and NLOS components of the Rician satellite-terrestrial channel, respectively. Additionally, $\pmb{\ell}_{k} \in \mathbb{C}^{M}$ represents the channel gain  \cite{9312147}.
Additionally, the UAV operates at a relatively low altitude, so a probabilistic channel model is adopted to characterize the likelihood of LoS and NLoS propagation for UAV–ground links. Consequently, the large-scale air-to-ground path gain for each user $k$ is modeled as a probabilistic mixture of LoS and NLoS components, given by
\begin{equation}
\tilde{\Lambda}_{k}= \eta_{k}\,\Lambda_{k}
  + \left(1-\eta_{k}\right)\,  \Lambda'_{k},
\end{equation}
where $\eta_{k}$ is the LoS probability and $\Lambda_{k}$, $\Lambda'_{k}$ are the deterministic path-gain terms under the two propagation conditions\cite{10681275}. Based on the 3GPP model, the LoS probability between the UAV and the user $k$ is 
\begin{equation}
    \eta_{k}=
\kappa_1- \kappa_2 \exp\left(- \kappa_3\arctan\frac{z^{(uav)}}{d_{k} \cos\theta_{k}^{(uav)}}\right),
\end{equation}
where $d_{k} = \|\mathbf{p}^{(uav)} - \mathbf{u}_{k}\|_2$ is the distance between UAV and user $k$, $z^{(uav)}$ is the fixed altitude of the UAV, and $\kappa_1, \kappa_2, \kappa_3$ lead to different propagation scenarios \cite{10681275}. Therefore, the channel vector between the UAV and user~$k$ is given by
\begin{equation}
    \mathbf{g}_{k} = \sqrt{\tilde{\Lambda}_{k}} e^{j2\pi f_{k} \frac{d_{k}}{c}} \mathbf{a}_{k}^{(uav)},
\end{equation}
where $\mathbf{a}_{k}^{(uav)}$ denotes the corresponding steering vector, $f_{k}$ is the subcarrier frequency, and $c$ is the speed of light.
\subsubsection{Communication Model for Rural Users}
\label{sec:Sat Com}
Regarding extra-urban communication, each user $k_1 \in \mathcal{Q}^{(sat)}$ is supported via satellite links, where dual-function signals are delivered through downlink satellite-terrestrial channels. At time index t, the dual-function signal of the satellite can be shown as
\begin{equation}
\mathbf{x}^{(sat)}(t) = \sum\nolimits_{k_1 \in \mathcal{Q}^{(sat)}} \mathbf{w}_{k_1} s^{\text{comm}}_{k_1}(t) + \sum\nolimits_{l_1 \in \mathcal{T}^{(sat)}}\mathbf{r}_{l_1}s^{\text{sens}}_{l_1}(t),
\end{equation}
where $\mathbf{w}_{k_1} \in \mathbb{C}^{M^{(sat)}}$ denotes the satellite transmit communication beamforming vector used to carry the communication data symbol, which is assumed to be zero-mean and of unit power
 $s^{\text{comm}}_{k_1}(t)$ of users~$k_1$. Similarly, $\mathbf{r}_{l_1} \in \mathbb{C}^{M^{(sat)} }$ denotes the transmit sensing beamforming vector at the LEO satellite to enable full degrees of freedom for target~$l_1$ with sensing symbol $s^{\text{sens}}_{l_1}$. 
To simplify the notation, the time index is omitted. The received signal at user $k_1$, denoted by $y_{k_1} \in \mathbb{C}$, is formulated as
\begin{align}\label{eq:y_com_SAT_longform}
    y_{k_1} &= {\mathbf{h}}_{k_1}^H {\mathbf{w}}_{k_1} s_{k_1}^{\text{comm}}
    + \sum\nolimits_{\substack{k_1' \in \mathcal{Q}^{{(sat)}} \\ k_1' \neq k_1}} {\mathbf{h}}_{k_1}^H {\mathbf{w}}_{k_1'} s^{\text{comm}}_{k_1'} \nonumber\\ &
    + \sum\nolimits_{l_1 \in \mathcal{T}^{(sat)}} \mathbf{h}_{k_1}^H \mathbf{r}_{l_1} s^{\text{sens}}_{l_1} +\sum\nolimits_{k_2 \in \mathcal{Q}^{(uav)}} \mathbf{g}_{k_1}^H\mathbf{v}_{k_2}s^{\text{comm}}_{k_2}\nonumber \\& +\sum\nolimits_{l_2 \in \mathcal{T}^{(uav)}} \mathbf{g}_{k_1}^H\mathbf{q}_{l_2}s^{\text{sens}}_{l_2}+ n_{k_1},
\end{align}
where $n_{k_1}$ is an additive white Gaussian noise (AWGN) process with zero mean and variance $\sigma^2$, i.e., $n_{k_1} \sim \mathcal{CN}(0, \sigma^2)$. To counteract this noise and Doppler, the receiver at each user applies joint delay synchronization and Doppler compensation using established methods, such as geometric-based estimation, Kalman filtering, or maximum likelihood techniques \cite{9628071}. These corrections ensure that the residual Doppler shift can be effectively mitigated, thus maintaining the integrity of the received signal. 
Therefore, the satellite's total transmit power is calculated by
$ P^{(sat)} = \sum_{k_1 \in \mathcal{Q}^{(sat)}} \|\mathbf{w}_{k_1}\|^2 + \sum_{l_1 \in \mathcal{T}^{(sat)}}\|\mathbf{r}_{l_1}\|^2.
$ In this expression, the first term accounts for the total transmit power allocated to satellite communication users via beamforming vectors $\mathbf{w}_{k_1}$, while the second term captures the power for sensing the targets through sensing beams $\mathbf{r}_{l_1}$.

\subsubsection{Communication Model for Urban Users}\label{sec:UAV com}
In an urban scenario, ground users $k_2 \in \mathcal{Q}^{(uav)}$ are served by a UAV hovering at a fixed altitude. Both the communication and sensing channels are assumed to be quasi-static and can be estimated through pilot signals.
We denote the aggregate dual-functional signal transmitted at the UAV as follows
\begin{equation}
\mathbf{x}^{(uav)} = \sum\nolimits_{k_2 \in \mathcal{Q}^{(uav)}} \mathbf{v}_{k_2} s^{\text{comm}}_{k_2} + \sum\nolimits_{l_2 \in \mathcal{T}^{(uav)} }{\bf q}_{l_2} s^{\text{sens}}_{l_2},
\end{equation}
where $s^{\text{comm}}_{k_2}$ is the communication symbol, $s^{\text{sens}}_{l_2}$ is the dedicated sensing signal and uncorrelated with $s^{\text{comm}}_{k_2}$. Additionally, $\mathbf{v}_{k_2}$ and $\mathbf{q}_{l_2}$ are beamforming vectors for communication and sensing tasks, respectively.
Furthermore, the transmit power of the UAV, which is calculated as follows
$P^{(uav)} = \sum_{k_2 \in \mathcal{Q}^{(uav)}} \|\mathbf{v}_{k_2}\|^2 + \sum_{l_2 \in \mathcal{T}^{(uav)}}\|\mathbf{q}_{l_2}\|^2.
$ Similarly, for user~$k_2 \in \mathcal{Q}^{(uav)}$, the received signal from the UAV can be expressed as
\begin{align}\label{eq:y_com_UAV_longform}
            y_{k_2} 
&={{{\bf g}_{k_2}^H {\bf v}_{k_2} s_{k_2}^{\text{comm}}}}+ {\sum\nolimits_{\substack{k_2' \in \mathcal{Q}^{(uav)}\\k_2' \neq k_2}}  {\bf g}_{k_2}^H {\bf v}_{k'_2} s^{\text{comm}}_{k'_2}} \nonumber \\
    		& + {\sum\nolimits_{l_2\in \mathcal{T}^{(uav)}} {\bf g}_{k_2}^H{\bf q}_{l_2} s_{l_2}^{\text{sens}}} +{\sum\nolimits_{k_1 \in \mathcal{Q}^{(sat)}} \mathbf{h}_{k_2}^H\mathbf{w}_{k_1}s^{\text{comm}}_{k_1}}\nonumber\\&+{\sum\nolimits_{l_1 \in \mathcal{T}^{(sat)}} \mathbf{h}_{k_2}^H\mathbf{r}_{l_1}s^{\text{comm}}_{l_1}}+  n_{k_2}, 
        \end{align}
where $n_{k_2}$ is the AWGN at user $k_2$ with $n_{k_2} \sim \mathcal{CN}(0,\sigma^2)$. 
\subsubsection{Downlink Transmission Rate}

The signal-to-interference-plus-noise ratio (SINR) of user $k^{(\xi)}$, served by platform $\xi \in \{\mathrm{sat},\mathrm{uav}\}$ are given by
\begin{equation}
\mathsf{SINR}_{k}^{(\xi)} = {\mathrm{DI}_{k}^{(\xi)}}/({\mathrm{SI}_{k}^{(\xi)} + \sigma^{2}}),
\end{equation}
where $\sigma^{2}$ is the noise variance and $B_u$~[MHz] denotes the system bandwidth. The desired signal power is expressed as one of the following two cases
\begin{equation}
\mathrm{DI}_{k_1}^{(\mathrm{sat})} = \bigl| \mathbf{h}_{k_1}^H \mathbf{w}_{k_1} \bigr|^2,
\qquad
\mathrm{DI}_{k_2}^{(\mathrm{uav})} = \bigl| \mathbf{g}_{k_2}^H \mathbf{v}_{k_2} \bigr|^2 .
\end{equation}
And the interference power for a satellite-served user $k_1$ and a UAV-served user $k_2$ is respectively given by
\begin{align}
\mathrm{SI}_{k_1}^{({sat})}
&= \sum\nolimits_{\substack{k_1' \in \mathcal{Q}^{({sat})}\\ k_1' \neq k_1}}
\! \bigl| \mathbf{h}_{k_1}^H \mathbf{w}_{k_1'} \bigr|^2
+ \sum\nolimits_{l_1 \in \mathcal{T}^{({sat})}}
\! \bigl| \mathbf{h}_{k_1}^H \mathbf{r}_{l_1} \bigr|^2 \notag \\
&\quad
+ \sum\nolimits_{k_2 \in \mathcal{Q}^{({uav})}}
\! \bigl| \mathbf{g}_{k_1}^H \mathbf{v}_{k_2} \bigr|^2
+ \sum\nolimits_{l_2 \in \mathcal{T}^{({uav})}}
\! \bigl| \mathbf{g}_{k_1}^H \mathbf{q}_{l_2} \bigr|^2 ,
\\[1mm]
\mathrm{SI}_{k_2}^{({uav})}
&= \sum\nolimits_{\substack{k_2' \in \mathcal{Q}^{({uav})}\\ k_2' \neq k_2}}
\! \bigl| \mathbf{g}_{k_2}^H \mathbf{v}_{k_2'} \bigr|^2
+ \sum\nolimits_{l_2 \in \mathcal{T}^{(uav)}}
\! \bigl| \mathbf{g}_{k_2}^H \mathbf{q}_{l_2} \bigr|^2 \notag \\
&\quad
+ \sum\nolimits_{k_1 \in \mathcal{Q}^{({sat})}}
\! \bigl| \mathbf{h}_{k_2}^H \mathbf{w}_{k_1} \bigr|^2
+ \sum\nolimits_{l_1 \in \mathcal{T}^{({sat})}}
\! \bigl| \mathbf{h}_{k_2}^H \mathbf{r}_{l_1} \bigr|^2 .
\end{align}
Hence, the corresponding downlink rate is obtained as
\begin{equation}
R_{k}^{(\xi)} = B_u \log_2\!\left(1+\mathsf{SINR}_{k}^{(\xi)}\right),
\end{equation}
which characterizes the achievable spectral efficiency of user $k$ under the considered satellite-UAV integrated transmission.

\subsection{Multi-target Sensing Network}
\label{sec:sensing}
For multi-target localization, ISAC-equipped aerial platforms (e.g., UAVs or satellites) transmit dual-functional waveforms toward ground or aerial targets. Reliable sensing hinges on designing transmit signals that support high-resolution detection and robust parameter estimation. A fundamental metric that characterizes the sensing capability is the transmit signal covariance matrix~\cite{9761984}, which captures the spatial power distribution and dictates the resulting sensing beampattern. It is defined as 
{\begin{equation}
    {\bf R}^{(\xi)} = \mathbb{E}\left\{{\bf x}^{(\xi)}\left({\bf x}^{(\xi)}\right)^H\right\}, (\xi) \in \{{uav, sat}\}. 
\end{equation}
With UAV and satellite platforms, the sensing goal is to maximize the power directed toward the target region, typically formulated as maximizing the transmit beampattern gain to enhance detection and parameter estimation. Since the beampattern is governed by the transmit covariance matrix, its design must focus energy toward the target while suppressing undesired directions. Formally, the beampattern gain at target $l \in \mathcal{T}^{(sat)} \cup \mathcal{T}^{(uav)}$ is expressed as 
$\mathrm{BG}^{(\xi)}_l = \mathbb{E}\left\{ (\mathbf{a}^{(\xi)}_l)^H\mathbf{R}^{(\xi)} \mathbf{a}^{(\xi)}_l\right\}.$
The signal elements are independent, and each has a zero mean.

\begin{remark}
With its high-altitude viewpoint, the satellite offers broad-area sensing, while the low-flying UAV provides fine-resolution coverage. Using prior target-location information \cite{10602493,9761984}, both platforms adapt their transmit power to focus on selected regions. This complementary sensing strategy leverages the satellite’s wide footprint and the UAV’s spatial precision to enhance overall sensing accuracy.
\end{remark}

\section{Joint Optimization of Communication and Sensing Energy Efficiency}
\label{sec:Joint_SAT_UAV}
The section formulates a unified optimization framework that maximizes system throughput while jointly allocating communication and sensing resources under QoS demands, power constraints, and sensing reliability requirements.
\subsection{Problem Formulation}
\label{sec:Problem_fomu}
Effective optimization of overall system performance is achieved by maximizing EE in both communication and sensing tasks. Accordingly, the EE optimization problem is formulated as follows:
\begin{subequations}\label{convert_problem}
\begin{align}
  \max_{\mathbf{W}, \tilde{\mathbf{R}}, \mathbf{V}, \mathbf{Q}}\;
       & \mathrm{EE} = \frac{\sum_{k_1 \in \mathcal{Q}^{(sat)}} R_{k_1} + \sum_{k_2 \in \mathcal{Q}^{(uav)}} R_{k_2}}{P^{(sat)} + P^{(uav)}}, \\
  \text{s.t.}\quad 
      & 0< P^{(\xi)} \leq P^{(\xi)}_{\max}, 
           \xi \in \{sat,uav\}, \label{eq:power_condition} \\
      &  R_{k^{(\xi)}}  \geq R_0^{(\xi)}, \xi \in \{sat,uav\}, \label{eq:comm_condition_1} \\
     &{\text{BG}_l^{(\xi)}} \geq \Gamma^{(\xi)}_0,  l\in \mathcal{T}^{(sat)} \cup \mathcal{T}^{(uav)},  \xi \in \{sat, uav\}. \label{eq:beampatter_gain conditions}
\end{align}
\end{subequations}
Here, $\mathbf{W}=\{\mathbf{w}_{k_1}\}$ and $\tilde{\mathbf{R}}=\{\mathbf{r}_{k_1}\}, \forall k_1 \in \mathcal{Q}^{(sat)}$, denote the satellite beamforming vectors for communication and sensing. Likewise, $\mathbf{V}=\{\mathbf{v}_{k_2}\}$ and $\mathbf{Q}=\{\mathbf{q}_{k_2}\}, \forall k_2 \in \mathcal{Q}^{(uav)}$, represent the UAV beamforming vectors. The parameters $P^{(\xi)}_{\mathrm{max}}, \forall \xi \in \{sat,uav\}$, specify the maximum transmit power, while $R_0^{(\xi)}$ is the minimum user rate requirement and $\Gamma_0$ is the required beampattern gain for sensing. Constraint \eqref{eq:power_condition} limits the transmit power of both platforms, \eqref{eq:comm_condition_1} enforces the QoS constraints for communication, and \eqref{eq:beampatter_gain conditions} guarantees sufficient beampattern gain for reliable sensing.

\subsection{Alternating Optimization}\label{sec:AO}
To efficiently address the fractional energy-efficiency maximization problem, we employ Dinkelbach’s iterative method. Let $\boldsymbol{\Upsilon} = \{\mathbf{W}, \tilde{\mathbf{R}}, \mathbf{V}, \mathbf{Q}\}$ denote the collection of all beamforming variables and optimization variables. We also introduce an auxiliary scalar parameter $\Psi$ to convert the fractional objective into a subtractive parametric form, enabling tractable optimization. Accordingly, the original fractional problem can be reformulated as
\begin{subequations}
\begin{align}
\label{eq:optimization_2}
    & \max_{\mathbf \Upsilon} f_1({\bf \Upsilon}) - \Psi f_2(\bf \Upsilon) \\
    & \text{s.t.}\quad  (\ref{eq:power_condition}),(\ref{eq:comm_condition_1}), (\ref{eq:beampatter_gain conditions}).
\end{align}
\end{subequations}
where $f_1(\mathbf{\Upsilon})$ and $f_2(\mathbf{\Upsilon})$
 are the aggregate throughput and total power, which are defined as follows
\begin{align}
   f_1(\boldsymbol{\Upsilon}) &=
      \sum_{k_1 \in \mathcal{Q}^{(sat)}} R_{k_1} + \sum_{k_2 \in \mathcal{Q}^{(uav)}} R_{k_2},\\
  f_2(\boldsymbol{\Upsilon}) &= P^{(sat)} + P^{(uav)}.
\end{align}
The aggregate throughput is the sum of achievable data rates over all satellite and UAV communication links under the current allocation $\boldsymbol{\Upsilon}$, while the total power consumption reflects the combined transmit power of both platforms based on their beamforming and control variables. 

Due to the similarity in signal models, interference patterns, and beamforming structures of the satellite and UAV platforms, we first derive the constraint transformations for the satellite case, while noting that the UAV case follows analogously.
Let $\boldsymbol{\Upsilon}^{(i)}$ and $\Psi^{(i)}$ denote the beamforming vectors and the auxiliary parameter, respectively, at the $i$-th iteration. The Dinkelbach update for the energy efficiency parameter at iteration $i+1$ is expressed as follows
\begin{equation}
\Psi^{(i+1)} = {f_1(\boldsymbol{\Upsilon}^{(i)})}/{f_2(\boldsymbol{\Upsilon}^{(i)})}.
\end{equation}
To obtain a Disciplined Convex Programming (DCP)-compliant reformulation of the sum–log utility, we introduce the lifted matrices $\overline{\bf W}_{k_1} = {\bf w}_{k_1}{\bf w}_{k_1}^H, \overline{\bf R}_{l_1} = {\bf r}_{l_1}{\bf r}_{l_1}^H, \overline{\bf V}_{k_2} = {\bf v}_{k_2}{\bf v}_{k_2}^H $,  $\overline{\bf Q}_{l_2} = {\bf q}_{l_2}{\bf q}_{l_2}^H$. In addition, we define non-negative auxiliary variables  $t^{(sat)}_{k_1}\triangleq\sqrt{\operatorname{Tr}(\overline{\mathbf W}_{k_1}^{ H}\mathbf{h}_{k_1}\mathbf{h}_{k_1}^{H})}$ and $t^{(uav)}_{k_2}\triangleq\sqrt{\operatorname{Tr}(\overline{\mathbf V}_{k_2}^{H}\mathbf{g}_{k_2}\mathbf{g}_{k_2}^{H})}$ for $k_1\in\mathcal Q^{(sat)}$ and $k_2\in\mathcal Q^{(uav)}$. 
These definitions lead to the following second-order cone (SOC) constraints
\begin{align}
\left(t^{(sat)}_{k_1}\right)^2 \le \operatorname{Tr}(\overline{\mathbf W}_{k_1}^{ H}\mathbf{h}_{k_1}\mathbf{h}_{k_1}^{H}),
\label{variable_t2}
\left(t^{(uav)}_{k_2}\right)^2 \le \operatorname{Tr}(\overline{\mathbf V}_{k_2}^{H}\mathbf{g}_{k_2}\mathbf{g}_{k_2}^{H}),
\end{align}
which are convex and compatible with DCP rules.
Let \(\boldsymbol\zeta=\{\zeta^{(sat)}_{k_1},\zeta^{(uav)}_{k_2}\}\) be a set of auxiliary variables.  By invoking the quadratic transform technique from \cite{10445319}, the utility function $f_1$, $\xi \in \{\mathrm{sat},\mathrm{uav}\} $, is reformulated as 
\begin{equation}
\label{eq:Quadratic-transform}
    \tilde{f_1}(\Upsilon)=\max \sum_{k_i\in\mathcal Q^{(\xi)}}\!\log\left(1+2\zeta^{(\xi)}_{k_i} t^{(\xi)}_{k_i}-\left(\zeta^{(\xi)}_{k_i}\right)^2 I^{(\xi)}_{k_i}\right),
\end{equation}
where $i \in \{1;2\}$ and $I_k^{(\xi)}$ denotes the interference generated jointly by the communication and sensing signals of both the UAV and satellite, is given by
\begin{align}
&I^{(\xi)}_{k} =
\sum_{\substack{i \in \mathcal{Q}^{(\xi)} \\ i \neq k}}
\mathrm{Tr}\!\left(\overline{\mathbf{W}}_{i}^{H}\overline{\mathbf{H}}^{(\xi)}_{k}\right)
+
\sum_{l \in \mathcal{T}^{(\xi)}}
\mathrm{Tr}\!\left(\overline{\mathbf{R}}_{l}^{H}\overline{\mathbf{H}}^{(\xi)}_{k}\right)\nonumber\\
&+
\sum_{\xi' \neq \xi}
\left(
\sum_{i \in \mathcal{Q}^{(\xi')}} \mathrm{Tr}\!\left(\overline{\mathbf{V}}_{i}^{H}\overline{\mathbf{H}}^{(\xi)}_{k}\right)
+
\sum_{l \in \mathcal{T}^{(\xi')}} \mathrm{Tr}\!\left(\overline{\mathbf{Q}}_{l}^{H}\overline{\mathbf{H}}^{(\xi)}_{k}\right)
\right)
+ \sigma^2,
\label{eq:Unified_I}
\end{align}
and the white noise affecting the received signal of user $k$ at platform $\xi$, $\overline{\mathbf{H}}^{(\xi)}_{k}=\mathbf h_{k^{(\xi)}} \mathbf{h}_{k^{(\xi)}}^H,$ and $\overline{\mathbf{G}}^{(\xi)}_{k}=\mathbf g_{k^{(\xi)}} \mathbf{g}_{k^{(\xi)}}^H$. Each auxiliary variable in $\pmb{\zeta}$ admits a closed-form maximizer and can be updated in each iteration according to the following expression
\begin{equation}
    \begin{aligned}
    \label{update_zeta}
        \zeta^{(sat)}_{k_1} = {t^{(sat)}_{k_1}}/{I^{(sat)}_{k_1}},~ \zeta^{(uav)}_{k_2} = {t^{(uav)}_{k_2}}/{I^{(uav)}_{k_2}}. \\
    \end{aligned}
\end{equation}
Therefore, \eqref{eq:Quadratic-transform} is jointly concave with respect to the beamforming variables. The communication QoS constraints with satellite in \eqref{eq:comm_condition_1}  can be equivalently reformulated as linear matrix inequalities (LMIs), $ \forall k_1 \in \mathcal{Q}^{(sat)}$, as
\begin{align}
\label{25d_convert}
\mathrm{Tr}\!\bigl(\mathbf{H}^{(sat)}_{k_1} \overline{\mathbf W}_{k_1}\bigr)- &\gamma^{(sat)}I_{k_1}^{(sat)}     \geq \sigma^{2} \gamma^{(sat)},
\end{align}
where  $\gamma^{(\xi)} = 2^{R_0^{(\xi)}} - 1$. Additionally, the total power constraint \eqref{eq:power_condition} can be expressed as
\begin{align}
   \sum\nolimits_{k_1 \in \mathcal{Q}^{(sat)}}\text{Tr}(\overline{\mathbf W}_{k_1})+\sum\nolimits_{l_1 \in {\mathcal{T}^{(sat)}}}\text{Tr}(\overline{\mathbf R}_{l_1}\bigr)\le P^{(sat)}_{\text{max}} \label{Power_sat_convert}
\end{align}
The beam pattern gain constraints are similarly reformulated as quadratic forms in terms of the lifted variables. Specifically, constraint (\ref{eq:beampatter_gain conditions}) with satellite platform can be rewritten as
\begin{align}
       \left({\bf a}^{(sat)}_{l_1}\right)^H\left(\sum_{k_1 \in \mathcal{Q}^{(sat)}}\overline{\bf W}_{k_1}+ \sum_{l_1' \in \mathcal{T}^{(sat)}}\overline{\bf R}_{l_1'} \right){\bf a}^{(sat)}_{l_1} \geq \Gamma^{(sat)}_0,\label{re_beampatter_gain_condition_1}
\end{align}
where $l_1 \in {\mathcal{T}^{(sat)}}$ and a similar transformation approach can also be applied to the UAV platform.
Then, the inner loop, which solves beamforming vectors, is addressed via sequential
convex programming, while the outer Dinkelbach iteration monotonically
updates \(\Psi\) until convergence, thereby delivering the global optimum
of the original fractional problem.
Let $ \widetilde{\bf \Upsilon} = \left\{\overline{\mathbf{W}}_{k_1}, \overline{\bf R}_{l_1}, \overline{\bf V}_{k_2}, \overline{\bf Q}_{l_2}, \zeta^{(sat)}_{k_1}, \zeta^{(uav)}_{k_2}, t^{(sat)}_{k_1}, t^{(uav)}_{k_2}\right\} $, with $ k_1 \in \mathcal{Q}^{(sat)}, l_1\in \mathcal{T}^{(sat)}, k_2 \in \mathcal{Q}^{(uav)}, l_2 \in \mathcal{T}^{(uav)}$.
Therefore, we reformulate the problem \eqref{eq:optimization_2}  as follows
\begin{subequations}
\begin{align}
\label{final_problem}
&\quad \max_{\tilde{\bf \Upsilon}}
\tilde{f}_1(\widetilde{\bf \Upsilon}) - \Psi f_2(\widetilde{\bf \Upsilon}), \\
&\text{s.t.} (\ref{variable_t2})
 (\ref{25d_convert}), (\ref{Power_sat_convert}), (\ref{re_beampatter_gain_condition_1}),\\
 &\quad t^{(sat)}_{k_1} \geq 0, t^{(uav)}_{k_2} \geq 0, \forall k_1 \in \mathcal{Q}^{(sat)}, k_2 \in \mathcal{Q}^{(uav)}\\
&\quad \operatorname{rank}(\overline{\mathbf W}_{k_1})=
\operatorname{rank}(\overline{\mathbf R}_{l_1})= 1, \forall k_1 \in \mathcal{Q}^{(sat)}, l_1 \in \mathcal{T}^{(sat)}, \label{rank_constraint_1} \\
&\quad \operatorname{rank}(\overline{\mathbf V}_{k_2})=
  \operatorname{rank}(\overline{\mathbf Q}_{l_2})=1, \forall k_2 \in \mathcal{Q}^{(uav)}, l_2 \in \mathcal{T}^{(uav)}, \label{rank_constraint_2} \\
  &\quad \overline{\mathbf W}_{k_1}\succeq0,\overline{\mathbf{R}}_{l_1}\succeq0, \forall k_1 \in \mathcal{Q}^{(sat)}, l_1 \in \mathcal{T}^{(sat)},\\
    &\quad \overline{\mathbf V}_{k_2}\succeq0,\overline{\mathbf Q}_{l_2}\succeq0, \forall k_2 \in \mathcal{Q}^{(uav)}, l_2 \in \mathcal{T}^{(uav)}\
\end{align}
\end{subequations}
The constraints in (\ref{variable_t2}), (\ref{25d_convert}), and (\ref{Power_sat_convert}) apply to both the UAV and satellite configurations. When $\Psi$ is fixed, problem~(\ref{final_problem}) remains non-convex due to the rank constraints in (\ref{rank_constraint_1})–(\ref{rank_constraint_2}). To address this difficulty, we invoke semidefinite relaxation (SDR) by removing the rank-one restrictions, thereby converting the problem into a convex semidefinite program that can be efficiently solved using CVX.
In contrast to conventional SDR techniques, no Gaussian randomization step is required. As demonstrated in~\cite{9481926,9570143}, the KKT conditions guarantee that an optimal rank-one solution always exists for this class of optimization problems. Hence, the relaxed formulation is tight, and the original beamforming variables can be recovered directly from the SDR solution. The complete solution procedure is summarized in Algorithm~\ref{alg:convex}.

\begin{algorithm}[ht]
    \caption{Energy–Efficient ISAC Beamforming}
    \label{alg:convex}
\begin{algorithmic}[1] 
\STATE Initialize randomly $\tilde{\pmb{\Upsilon}}^{(0)}$;\label{step:init}\\
\STATE Compute $\Psi^{(0)}\leftarrow \tilde{f}_1(\tilde{\pmb{\Upsilon}}^{(0)})/f_2(\tilde{\pmb{\Upsilon}}^{(0)})$;\label{step:psi0}\\
\STATE  Update $\zeta^{(sat)}_{k_1}$ and $\zeta^{(uav)}_{k_2}$ based on ~\eqref{update_zeta};\label{step:zeta0}\\
\STATE $i\leftarrow 0$;\label{step:i0}\\

\WHILE{$|\tilde{f}_1(\tilde{\pmb{\Upsilon}}^{(i)})-\Psi^{(i)}f_2(\tilde{\pmb{\Upsilon}}^{(i)})|>\varepsilon$ \textbf{and} $i<G_{\max}$}
\label{step:while}
\STATE  Calculate $\tilde{\pmb{\Upsilon}}^{(i+1)}$ by solving the convexified sub-problem;\\
\STATE  $\Psi^{(i+1)}\leftarrow \tilde{f}_1(\tilde{\pmb{\Upsilon}}^{(i)})/f_2(\tilde{\pmb{\Upsilon}}^{(i)})$\label{step:psiupdate}, and $i\leftarrow i+1$;\label{step:iupdate}
\ENDWHILE

\end{algorithmic}
\end{algorithm}

\begin{figure*}[t]
    \centering
    \subfloat[]{%
        \includegraphics[width=0.24\textwidth]{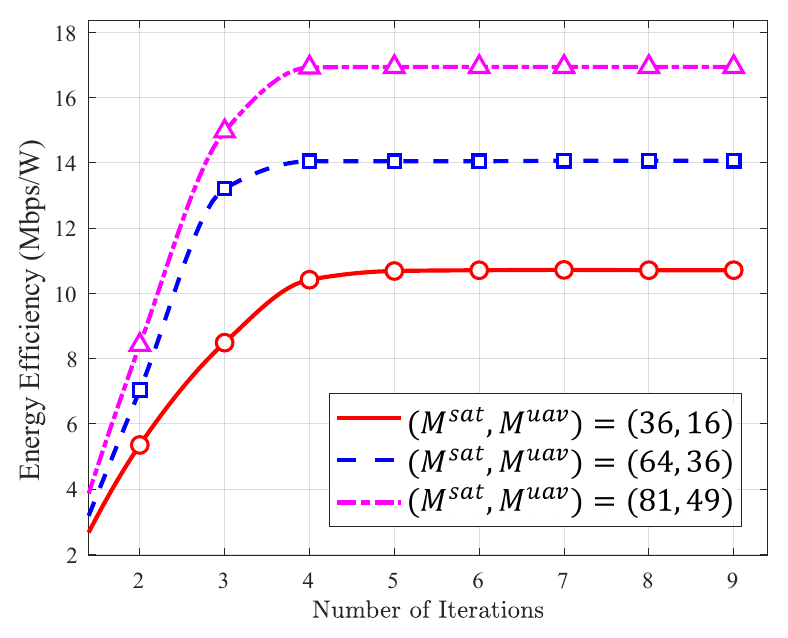}%
        \label{fig:convergence}}
    \hfill
    \subfloat[]{%
        \includegraphics[width=0.245\textwidth]{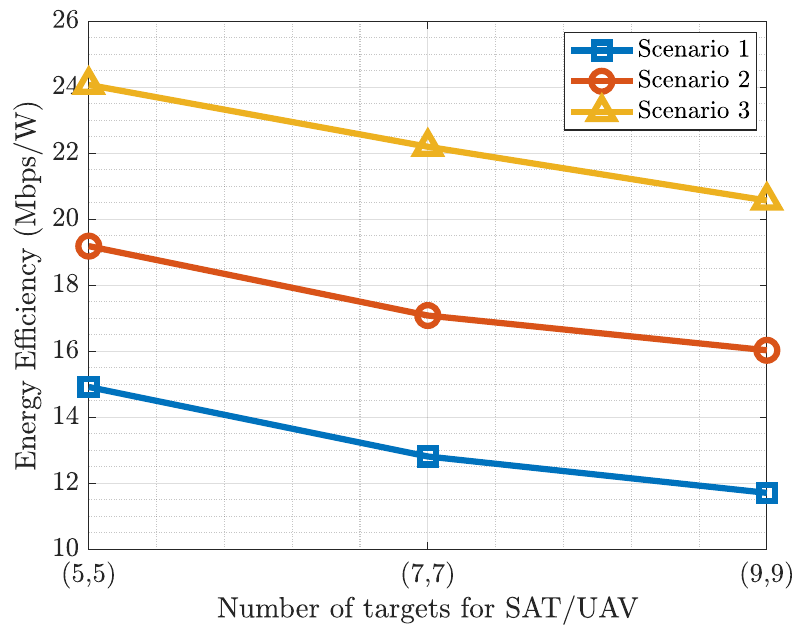}%
        \label{fig:increase_target}}
    \hfill
    \subfloat[]{%
        \includegraphics[trim={0 -1.2cm 0 0},clip,width=0.242\textwidth]{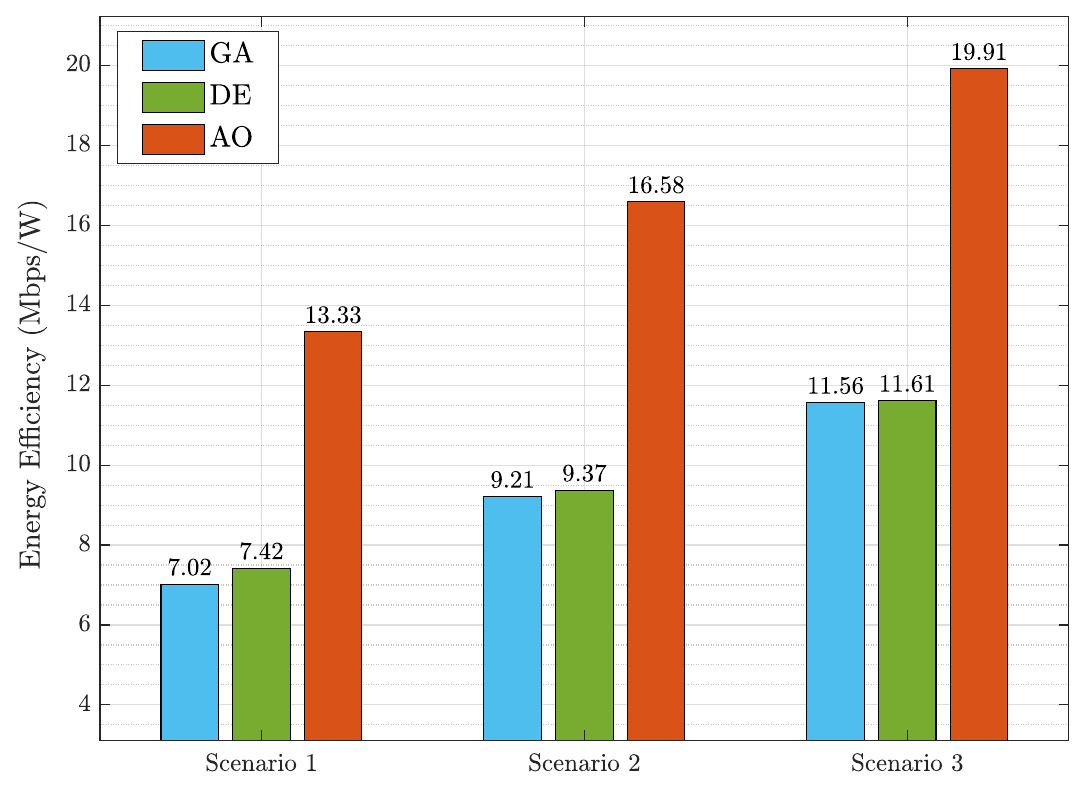}%
        \label{fig:compare_alg}}
    \hfill
    \subfloat[]{%
        \includegraphics[trim={0 0.5cm 0 0},clip,width=0.263\textwidth]{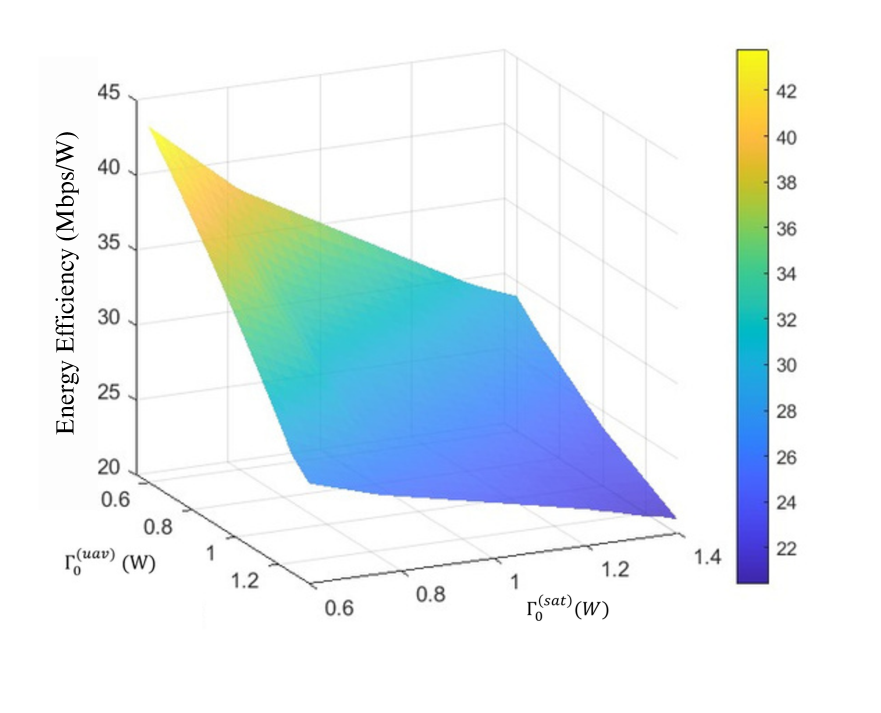}%
        \label{fig:beampattern}}
    \caption{Overall system performance across sensing and communication tasks:
    (a) Convergence behavior of Algorithm~\ref{alg:convex} over iterations; 
    (b) Impact of increasing the number of sensing targets on the achieved EE for various user densities; 
    (c) Comparison of EE achieved by proposed AO algorithms with other algorithms; 
    (d) EE performance as a function of the beampattern threshold.}
    \label{fig:combined_results}
\end{figure*}

Now we analyze the computation complexity, by $\texttt{dim}$ denotes the size of the largest semidefinite block.  
An interior-point method solves one SDP in 
$\mathcal{O}\!\bigl(\sqrt{\texttt{dim}}\,\texttt{dim}^{6}\log(1/\hat\varepsilon)\bigr)$ 
operations, where $\hat\varepsilon$ is the duality-gap tolerance.  
Including $I_{\mathrm{inner}}$ convexification steps and $I_{\mathrm{outer}}$ outer updates gives the overall cost
$
\mathcal{O}\!\Bigl(I_{\mathrm{outer}} I_{\mathrm{inner}} \sqrt{\texttt{dim}}\,\texttt{dim}^{6}\log(1/\hat\varepsilon)\Bigr).
$
Relaxing the SDP to an SOCP reduces the dominant cost from $\texttt{dim}^{6}$ to $\texttt{dim}^{3}$, yielding
$
\mathcal{O}\!\Bigl(I_{\mathrm{outer}} I_{\mathrm{inner}} \sqrt{\texttt{dim}}\,\texttt{dim}^{3}\log(1/\hat\varepsilon)\Bigr),
$
providing an approximate cubic speed-up while preserving the same iteration structure.

\section{Numerical Results}\label{sec:numerical results}
In this section, we present numerical results to evaluate the performance of the proposed algorithms. Satellite parameters are taken from practical Starlink constellation simulations \cite{9566290}, and UAV characteristics follow the 3GPP Release-20 models \cite{3gpp2025release20}. All experiments were executed on a high-performance workstation equipped with an AMD Ryzen 9 7950X with 16 cores 4.50 GHz, 128 GB RAM, running a 64-bit Windows OS. Key system parameters include: $c = 3\times10^8 \mathrm{m/s}$, Boltzmann constant $\kappa = 1.38\times10^{-23} \mathrm{J/K}$, and antenna spacing $d = \lambda/2$, with full details provided in TABLE.~\ref{table:simulation_parameters}. The scenario includes up to 35 users and as many as 20 spatial and terrestrial sensing targets to examine scalability under dense deployments.

\begin{table}[ht]
\centering
\caption{Simulation settings for satellite and UAV systems.}
\label{table:simulation_parameters}
\begin{tabular}{l|c|c}
\hline\cline{1-3}
\textbf{Parameter} & \textbf{LEO Satellite} & \textbf{UAV} \\
\hline\hline
Number of antennas & $\{36,64,81\}$ & $ \{16,36,49\}$ \\
\hline
Area covered & $1.5\mathrm{km}\times 1.5\mathrm{km}$ & $0.3\mathrm{km}\times 0.3\mathrm{km}$ \\
\hline
Signal frequency & $ 35\,\mathrm{GHz}$ & $ 28\,\mathrm{GHz}$ \\
\hline
Channel bandwidth & $ 20\,\mathrm{MHz}$ & $ 10\,\mathrm{MHz}$ \\
\hline
Distance to UEs & $ [550, 2700]\,\mathrm{km}$ & $  [100, 150]\,\mathrm{m}$ \\
\hline
Noise  & $-110\,\mathrm{dBm}$ & $ -110\,\mathrm{dBm}$ \\
\hline
Max transmit power & $6\,\mathrm{W}$ & $ 3\,\mathrm{W}$ \\\cline{1-3}
\hline\cline{1-3}
\end{tabular}
\end{table}

We investigate the convergence time of the proposed AO algorithm under three different user population scenarios, as depicted in Fig.~\ref{fig:convergence}. It is shown that the Alg.\ref{alg:convex} consistently achieves a fast convergence rate reaching the optimal value within 10 iterations on average. 
Moreover, we fix the number of users in each scenario and increase the number of targets to evaluate the network performance under varying interference levels.
Fig.\ref{fig:increase_target} illustrates the relationship between the number of targets and the system’s EE. As the number of targets increases, the EE gradually decreases. This behavior can be explained as with more targets, the system must allocate its transmission resources, such as power and bandwidth, across multiple spatial directions. This leads to higher energy consumption per successfully transmitted data unit. Moreover, increasing the number of targets intensifies mutual interference and reduces the spatial gains typically achieved in MIMO systems. Consequently, the overall throughput per unit of consumed power declines, resulting in reduced EE.


Furthermore, to evaluate the effectiveness of Alg.~\ref{alg:convex}, we compare it with two benchmark algorithms, namely Genetic Algorithm (GA) \cite{11134755} and Differential Evolution (DE) \cite{10721339}, under identical system settings. As shown in Fig.~\ref{fig:compare_alg}, Alg.~\ref{alg:convex} consistently achieves superior energy efficiency across all scenarios. Specifically, in Scenario~1, it improves EE by approximately 89.9\% and 79.6\% compared to GA and DE, respectively. The corresponding gains in Scenario~2 are about 80.0\% and 76.9\%, while in Scenario~3, the improvements reach 72.2\% over GA and 71.4\% over DE.
Finally, we examine the impact of the beam pattern threshold requirements of the satellite and UAV, denoted by $\Gamma_0^{(\mathrm{sat})}$ and $\Gamma_0^{(\mathrm{uav})}$. As illustrated in Fig.~\ref{fig:beampattern}, the system EE decreases as either threshold increases. This behavior arises because higher sensing power requirements demand greater transmit power allocation, increasing energy consumption and reducing the achievable communication rate, which ultimately degrades the overall EE.

\section{Conclusion}\label{sec:conclusion}
This paper investigates EE maximization in satellite–UAV assisted ISAC systems, jointly optimizing communication and sensing under realistic channels. An AO-based algorithm is developed to handle the beamforming, power, QoS, and beampattern constraints, achieving fast convergence to a stationary solution. Simulations show that the proposed framework delivers high EE while maintaining robust sensing and communication performance across diverse scenarios, demonstrating the potential of integrating satellite and UAV platforms in NsG networks. Future extensions may explore multi-satellite/UAV architectures and learning-driven resource allocation.

\bibliographystyle{IEEEtran}
\bibliography{References} 
\end{document}